\documentclass[12pt]{article}
\usepackage{rotating}

\begin{document}

\vspace*{2.5cm}

\begin{center}
{\Large {\sc Group Theoretical Properties and Band Structure of the Lam\'{e}
Hamiltonian}}

\vspace{1.8cm}

Hui LI\footnote{%
email: huili@nst4.physics.yale.edu}, Dimitri KUSNEZOV\footnote{%
email: dimitri@nst4.physics.yale.edu}, Francesco IACHELLO

{\sl Center for Theoretical Physics, Sloane Physics Laboratory,\\[0pt]
Yale University, New Haven, CT 06520-8120}

\vskip 1.2 cm

{\it December 1999}

\vspace{1.2cm}

\parbox{13.0cm}
{\begin{center}\large\sc ABSTRACT \end{center}
{\hspace*{0.3cm}
We study the group theoretical properties of the Lam\'e equation
and its relation to $su(1,1)$ and $su(2)$. We compute the band
structure, dispersion relation and transfer matrix and discuss the 
dynamical symmetry limits.}}
\end{center}

\vspace{3mm}

\noindent PACs numbers: 03.65.Fd, 02.20.-a, 02.20.Sv, 11.30.-j\newline

\noindent keywords: representation theory, dynamical symmetry, exactly
solvable models, periodic potentials, band structure.

\newpage

\setcounter{page}{2} 

\section{Introduction}

The use of group theory to study bound and scattering states has found wide
application in physics\cite{intro}. A common starting point is to identify a
spectrum generating algebra or SGA for the problem of interest. This is
possible when the Hamiltonian can be expressed in terms of generators of
some algebra $G$. Representation theory then can be used to identify exactly
solvable limits of the theory, which can then be translated to explicit
forms of the Hamiltonian\cite{prldk}.  Connections between bound state or
scattering state problems and representations of compact and non-compact
groups are well understood. The remaining category, having to do with band
structure and periodic potentials, was suggested initially in
\cite{gursey}, but remained an open problem until recently. In  \cite{prllk}
it was shown that dynamical symmetry techniques and representation
theory can be used to solve band structure problems. In
particular,  for the Scarf Hamiltonian, 
band structure arises when one uses the complementary
series of the projective representations of $su(1,1)$ (and $so(2,2)$ for the
extended Scarf potential).  Further, these dynamical symmetries allowed
simple evaluation of the transfer matrix and dispersion relations. In the
present study, we consider the band structure problem associated with the
Lam\'e equation, which is not a dynamical symmetry situation, but more
generally that of a SGA. We discuss several realizations of the Lam\'e
equation and its relation to  $su(2)$ and $su(1,1)$, including a discussion
of   dynamical symmetry limits, generators, dispersion relation and transfer
matrix.

While the Lam\'e equation at first might seem to be an obscure differential
equation, it does find surprisingly wide application in physics. For
instance, it has been shown that if one uses the periodic potentials in
super--symmetric quantum mechanics, the super partners of the Lam\'{e}
potentials are distinctly different from the original except for $n=1$ which
is not self-isospectral. This provides new solvable periodic potentials\cite
{a}. The Lam\'{e} equation also appears in the topics ranging from solitons
to exactly-solvable models. This includes associations with solutions of the
periodic KdV equation\cite{novikov}, BPS monopoles\cite{sutcliffe},
sphaleron solutions of the (1+1)-dimensional abelian Higgs model\cite
{brihaye}, sine-Gordon solitons\cite{Liang}, as well as relations to
Calogero-Moser systems\cite{enolskii}. The analysis of this equation has
many group theoretical aspects as well. For instance, the Lam\'{e} equation
can be written in terms of the composition of two first-order matrix
operators, where the coefficients of which satisfy so(3) Nahm's equation\cite
{b}. It also arises naturally in the context of $su(2)$ when one tries to
separate Laplace's equation in certain coordinate systems\cite{miller,patera}
as well as in the general classification of Lie algebraic potentials\cite
{karman} and in quasi-exactly solvable $sl(2)$ models\cite{turbiner}. The
discussion of band structure in the context of group theory is more recent
however\cite{gursey}, although the system was not solved and the transfer
matrix was not determined.

The Jacobian form of the Lam\'e equation is
\begin{equation}
-\frac{d^{2}}{dx^{2}}+\kappa ^{2}\ell (\ell +1){\rm sn}^{2}x={\cal E}.
\end{equation}
We will view this as a Schr\"odinger equation with mass $M=1/2$ when $x
$ is real valued. This equation was first studied group
theoretically in \cite{patera}. The elliptic functions sn$\alpha =$sn$(\alpha |\kappa )$,
cn$\alpha ={\rm cn(}\alpha |\kappa )$, and ${\rm dn}\alpha ={\rm dn(}\alpha
|\kappa )$ are doubly-periodic functions in the complex plane, of modulus $
\kappa $, where $0\leq \kappa \leq 1$. We will omit the modulus except in
functions where it differs from $\kappa $. The complementary modulus is
defined as $\kappa ^{\prime }=(1-\kappa ^{2})^{1/2}$. The periods of the
Jacobi elliptic functions are related to the complete elliptic integrals $
K=(\pi /2)F(1/2,1/2,1;\kappa ^{2})$ and $K^{\prime }=(\pi
/2)F(1/2,1/2,1;\kappa ^{\prime 2})$, where $F$ is the hypergeometric
function, by:

\begin{equation}
\begin{array}[t]{ll}
{\rm sn}\,(\alpha +\tau )={\rm sn}\,\alpha \qquad  & \tau =2iK^{\prime
},\quad 4K+4iK^{\prime },\quad 4K \\ 
{\rm cn}\,(\alpha +\tau )={\rm cn}\,\alpha  & \tau =4iK^{\prime },\quad
2K+2iK^{\prime },\quad 4K \\ 
{\rm dn}\,(\alpha +\tau )={\rm dn}\,\alpha  & \tau =4iK^{\prime },\quad
4K+4iK^{\prime },\quad 2K
\end{array}
\end{equation}
In the limit $\kappa =1$, the real period related to $K$ becomes infinite,
and the functions are no longer periodic on the real axis. \ Along the real
axis, the potential ${\rm sn}^{2}x$ is periodic and bounded, while along the
imaginary axis, it has periodic singularities of the type $1/x^{2}.$
Examples  are shown in Fig. 1 for $\kappa ^{2}=1/2$ for potentials $V(x)=
{\rm sn}^{2}(ax+b)$, with several choices of $a$ and $b$.

The Lam\'e equation has also been related to the band structure problem
associated with the P\"oschl-Teller potential, $V_{pt}(x)\sim 1/\cosh ^{2}x$
\cite{sutherland,prllk},  by using the expansions for the elliptic
functions in terms of trigonometric functions found in the exercises of
Whittaker and Watson\cite{whitwat}. Specifically, one can start with the 1-d
band structure problem 
\begin{equation}
H=-\frac{d^{2}}{dx^{2}}+\sum_{k=-\infty }^{\infty }\frac{g}{\cosh
^{2}((x-ka)/x_{0})}.
\end{equation}
If we write the coupling constant as 
\begin{equation}
g=-\frac{\ell (\ell +1)}{x_{0}^{2}}
\end{equation}
we can make the coordinate transformation 
\begin{equation}
x=i\left( \frac{\pi x_{0}}{2K}\right) z-\left( \frac{a+i\pi x_{0}}{2}\right) 
\end{equation}
which  transforms the Hamiltonian to the Lam\'e equation. In this article, we
first discuss the relation of $su(2)$ Hamiltonians to the Lam\'e equation,
followed by a discussion of band edges. Next we consider $su(1,1)$ and the
relation to scattering states in the dynamical symmetry limits. 
Finally we derive the transfer matrix and dispersion relation for the Lam\'e
equation followed by the relation to the group theoretical SGA\ Hamiltonians.

\section{$su(2)$ Realizations of the Lam\'e Equation}

\subsection{Sphero-conal Coordinates for $su(2)$}

\bigskip

In order to realize the Lam\'e equation from $su(2)$\ or $su(1,1)$, we will
use coordinate systems defined in terms of Jacobi elliptic functions. These
are often refered to as sphero-conal\cite{Arscott} or conical\cite{miller}
coordinates. One can find variations that have been considered in the past,
and not all provide a good description for all values of the modulus $\kappa
.$ Consider first the mapping: 
\begin{equation}
x=\kappa \,{\rm sn}\,\alpha \,{\rm sn}\,\beta ,\,y=i\frac{\kappa }{\kappa
^{\prime }}\,{\rm cn}\,\alpha \,{\rm cn}\,\beta ,\,z=\frac{1}{\kappa
^{\prime }}\,{\rm dn}\,\alpha \,{\rm dn}\,\beta ,
\end{equation}
The ranges of the angles $\alpha ,\beta $ are defined in terms of the
elliptic integrals $K$ and $K^{\prime }$, so that they depend on the value
of $\kappa $. Specifically $-2K<\alpha <2K$ and $\,K\leq \beta
<K+2iK^{^{\prime }}$ where $\kappa ^{\prime 2}=1-\kappa ^{2}$, and $
x^{2}+y^{2}+z^{2}=1.$ The contours of constant $\alpha $ and $\beta $ cover
the sphere as illustrated in Fig. 2(a). For $\kappa =1/2$, four typical
contours (for $\alpha =\pm K/2$ and $\beta =K+iK^{\prime }/2,K+3iK^{\prime
}/2$)\ are shown in\ Fig. 2(b). This parametrization can be used to
construct the generators of $su(2)$. A direct computation of the generators
yields the realization (I) given in Table 1. These
generators satisfy the commutation relations in the form $\left[ L_{a},L_{b}
\right] =\epsilon _{abc}L_{c}$ with Casimir invariant
\begin{eqnarray}
C_{2} &=&L_{x}^{2}+L_{y}^{2}+L_{z}^{2}  \nonumber \\
&=&\frac{1}{\kappa ^{2}({\rm sn}^{2}\,\alpha -{\rm sn}\,^{2}\beta )}\left( 
\frac{\partial ^{2}}{\partial \alpha ^{2}}-\frac{\partial ^{2}}{\partial
\beta ^{2}}\right) .
\end{eqnarray}
The $su(2)$ algebras we discuss here have Casimir invariants which have
expectation value $-\ell (\ell +1)$. We now consider which types of
quadratic Hamiltonians of the form 
\begin{equation}
H=\sum_{j}\eta _{j}L_{j}^{2}
\end{equation}
can be constructed which result in Schr\"odinger equations in $\alpha $ and $
\beta .$ Specifically, when the coordinate representation of $H$ contains
differential opertors only of the type $\partial ^{2}/\partial \alpha ^{2}$
and $\partial ^{2}/\partial \beta ^{2}$, one can use $H$ together with $C_{2}
$ to perform a separation of variables resulting in two independent Lam\'e
equations. We find the three forms in the top right of Table 1, which are
simply related through the Casimir operator. To demonstrate the separation,
consider $H_{3}$, 
\begin{eqnarray}
H_{3} &=&L_{z}^{2}+\kappa ^{2}L_{y}^{2}  \nonumber \\
&=&\frac{1}{({\rm sn}^{2}\,\beta -{\rm sn}\,^{2}\alpha )}\left( {\rm sn}
^{2}\,\beta \frac{\partial ^{2}}{\partial \alpha ^{2}}-{\rm sn}^{2}\,\alpha 
\frac{\partial ^{2}}{\partial \beta ^{2}}\right) ,  \label{ham1}
\end{eqnarray}
and require it to be a constant of the motion with eigenvalue $H_{3}={\cal E}
$. Using basis states denoted by the direct product, $\Psi (\alpha ,\beta
)=\psi (\alpha )\phi (\beta )$, we arrive at two identical Lam\'{e}
equations with identical eigenvalue: 
\begin{eqnarray}
\ \left[ -\frac{d^{2}}{d\alpha ^{2}}+\kappa ^{2}\ell (\ell +1)\,{\rm sn}
^{2}\alpha \right] \psi (\alpha ) &=&{\cal E}\psi (\alpha ), \\
\ \left[ -\frac{d^{2}}{d\beta ^{2}}+\kappa ^{2}\ell (\ell +1)\,{\rm sn}
^{2}\beta \right] \phi (\beta ) &=&{\cal E}\phi (\beta ).  \nonumber
\end{eqnarray}
Thus solution of the eigenvalue problem for $H_{3}=L_{z}^{2}+\kappa
^{2}L_{y}^{2}$ yields the single valued solutions to these equations, which
will correspond to the band edges. One should keep in mind that while $
\alpha $ is defined on the real axis, $\beta $ is complex. Also, while the
Hamiltonians $H_{k}$ are well defined in their algebraic form, the
generators and coordinates have singularities in both $\kappa =0$ and $
\kappa =1$ limits. Hence we consider a different realization below.

To reduce these Hamiltonians to more conventional Schr\"odinger
equations in $\alpha$ and $\beta$, we would
like the coordinates to be defined along the real axis. To do so, we make a
shift (i) $\beta \rightarrow \beta +K+iK^{^{\prime }}$ followed by (ii) $
\beta \rightarrow -i\beta $. This results in the transformations: 
\begin{eqnarray}
{\rm sn}(\beta |\kappa ) &\rightarrow &\frac{1}{\kappa }{\rm dn}(\beta
|\kappa ^{\prime }),  \nonumber \\
{\rm cn}(\beta |\kappa ) &\rightarrow &\frac{\kappa ^{\prime }}{i\kappa }
{\rm cn}(\beta |\kappa ^{\prime }), \\
{\rm dn}(\beta |\kappa ) &\rightarrow &\kappa ^{\prime }{\rm sn}(\beta
|\kappa ^{\prime }).  \nonumber
\end{eqnarray}
The new coordinate system is no longer singular, and is parameterized as 
\begin{equation}
x={\rm sn}\,\alpha \,{\rm dn}(\beta |\kappa ^{\prime }),\,y={\rm cn}\,\alpha
\,{\rm cn}(\beta |\kappa ^{\prime }),\,z={\rm dn}\,\alpha \,{\rm sn}(\beta
|\kappa ^{\prime }),
\end{equation}
where $-2K<\alpha <2K,\,-K^{^{\prime }}\leq \beta <K^{^{\prime }}$ are now
both real valued. Contours of constant $\alpha $ and $\beta $ are shown in
Fig. 2(c) for $\alpha =\pm K/2$ and $\beta =\pm K^{\prime }/2$. If we
construct the generators of $su(2)$ in this basis, we obtain the realization
(II)\ given in Table 1. The Casimir invariant now has the form 
\[
C_{2}=L_{x}^{2}+L_{y}^{2}+L_{z}^{2}=\frac{-1}{\kappa ^{2}{\rm cn}^{2}\alpha
+\kappa ^{\prime 2}{\rm cn}^{2}(\beta |\kappa ^{\prime })}\left( \frac{
\partial ^{2}}{\partial \alpha ^{2}}+{\rm \ }\frac{\partial ^{2}}{\partial
\beta ^{2}}\right) 
\]
The quadratic Hamiltonians which give rise to Schr\"odinger equations when
combined with $C_{2}$ are given in the lower right of Table 1. Consider
again $H_{3}=L_{z}^{2}+\kappa ^{2}L_{y}^{2}$. If we set $H_{3}={\cal E}$ and 
$C_{2}=-\ell (\ell +1)$, we obtain 
\begin{eqnarray}
 \left[ -\frac{d^{2}}{d\alpha ^{2}}+\kappa ^{2}\ell (\ell +1){\rm sn}
^{2}\alpha \right] \psi (\alpha ) &=&{\cal E}\psi (\alpha ), \\
\ \left[ -\frac{d^{2}}{d\beta ^{2}}+\kappa ^{\prime 2}\ell (\ell +1){\rm sn}
^{2}(\beta |\kappa ^{\prime })\right] \phi (\beta ) &=&(\ell (\ell +1)-{\cal 
E})\phi (\beta ).
\end{eqnarray}
In this case the coordinates are both real valued. For integer $\ell $, and
both potentials are non-negative, \ we see that $0\leq {\cal E}\leq \ell
(\ell +1).$ Further the ground state of one Hamiltonian corresponds the the
highest energy state of the other, and vice-versa. (Similar equations
are obtained if we use $H_{2}$ or $H_{1}.$) Before we discuss some of the
general properties of these equations, we consider first the dynamical
symmetry limits which are when $\kappa =0$ or $1$.

\subsection{$\kappa =0$ Limit}

In the limit $\kappa =0$, and $H_{3}=L_{z}^{2}=$ $m^{2}$, so that ${\cal E}
=m^{2}$, and the coordinate Hamiltonians are: 
\begin{eqnarray}
\ -\frac{d^{2}}{d\alpha ^{2}}\psi (\alpha ) &=&{\cal E}\psi (\alpha ), \\
\ \left[ -\frac{d^{2}}{d\beta ^{2}}-\frac{\ell (\ell +1)}{\cosh ^{2}\beta }
\right] \phi (\beta ) &=&-{\cal E}\phi (\beta ).
\end{eqnarray}
The generators in this limit have the simpler form: 
\begin{eqnarray}
I_{\pm } &=&\ e^{\pm i\alpha }\left[ \mp \cosh \beta \ \frac{\partial }{
\partial \beta }+i\sinh \beta \frac{\partial }{\partial \alpha }\right]  
\nonumber \\
I_{3} &=&\ -i\frac{\partial }{\partial \alpha }
\end{eqnarray}
The first equation is free motion, while the second corresponds to bound
states of the P\"oschl-Teller potential. For the discrete representations of $
su(2)$, one obtains the bound state spectrum $\ -{\cal E}=-m^{2}.$ The
wavefunctions are of the form: 
\begin{equation}
\Psi (\alpha ,\beta ;\kappa =0)\sim P_{\ell }^{m}(\tanh \beta )\exp (\pm
im\alpha ),\qquad m=-\ell ,...,\ell ;\quad \ell =0,1,2,...
\end{equation}
with $-\pi <\alpha <\pi ,\,-\infty <\beta <\infty .$

\subsection{$\kappa =1$ Limit}

In the limit $\kappa =1$, $H_{3}=L^{2}-L_{x}^{2},$ the equations reduce to 
\begin{eqnarray}
\ \left[ -\frac{d^{2}}{d\alpha ^{2}}-\frac{\ell (\ell +1)}{\cosh ^{2}\alpha }
\right] \ \psi (\alpha ) &=&[{\cal E}-\ell (\ell +1)]\psi (\alpha )={\cal E}
^{\prime }\psi (\alpha ) \\
-\frac{d^{2}}{d\beta ^{2}}\phi (\beta ) &=&[\ell (\ell +1)-{\cal E}]\phi
(\beta )=-{\cal E}^{\prime }\phi (\beta )
\end{eqnarray}
The generators are the same as in the $\kappa =0$ limit, with $\alpha $ and $
\beta $ interchanged. The second equation is now free motion, while the
first corresponds to bound states of the P\"oschl-Teller potential with a
shifted eigenvalue. For the discrete representations of $su(2)$, one obtains
the bound state spectrum ${\cal E}^{\prime }=-m^{2}.$ The wavefunctions are
of the form: 
\begin{equation}
\Psi (\alpha ,\beta ;\kappa =1)\sim P_{\ell }^{m}(\tanh \alpha )\exp (\pm
im\beta ),\qquad m=-\ell ,...,\ell ;\quad \ell =0,1,2,...
\end{equation}
with $-\infty <\alpha <\infty ,\,-\pi <\beta <\pi .$ \ 

\subsection{ Band Edges and $su(2)$}

For general values of $\kappa $, the Lam\'e Hamiltonians are periodic (as seen
in Fig. 1), and will have band structure. As the discrete representations of 
$su(2)$ correspond to single-valued wavefunctions, the discrete
representations can at most describe the band edges. Since we have an
algebraic realization of the Hamiltonians whose spectrum corresponds to the
Lam\'e equations, we can obtain the eigenvalues of the Hamiltonian $H_{k}$
(and hence the band edges) by a direct diagonalization in the spherical
harmonic basis $\left| \ell m\right\rangle $, which can then be related to
the results for the elliptic basis through the coordinate transformations.
The resulting functions are 
doubly periodic solutions of Lam\'e's equation known as Lam\'e
polynomials\cite{Arscott}. These polynomials are of the form
sn$^a x$cn$^b x$dn$^c x F_p($sn$^2 x )$, where $a,b,c=0,1$ and
$a+b+c+2p=\ell$. Here $F_p(z)$ is a polynomial in $z$ of order
$p$.
A discussion of these functions can be found in \cite{bateman,whitwat,Arscott}.
For $\ell =1$, there are three eigenstates of $H_{3}$ given by 
\begin{eqnarray}
\Psi _{1}(\alpha ,\beta ;\kappa ) &=&\left| 10\right\rangle \sim {\rm sn}
\,\alpha {\rm dn}(\beta |\kappa ^{\prime })  \nonumber \\
\Psi _{2}(\alpha ,\beta ;\kappa ) &=&\frac{1}{\sqrt{2}}\left[ \left|
11\right\rangle +\left| 1-1\right\rangle \right] \sim {\rm \ cn}\,\alpha 
{\rm cn}(\beta |\kappa ^{\prime }) \\
\Psi _{3}(\alpha ,\beta ;\kappa ) &=&\frac{1}{\sqrt{2}}\left[ \left|
11\right\rangle -\left| 1-1\right\rangle \right] \sim {\rm \ dn}\,\alpha 
{\rm sn}(\beta |\kappa ^{\prime })  \nonumber
\end{eqnarray}
with eigenvalues $E_{1}=1+\kappa ^{2}$, $E_{2}=1$, $E_{3}=\kappa ^{2}$. In
the dynamical symmetry limits, these three states reduce to those discussed
above.

For $\ell =2$, we have 
\begin{eqnarray}
\Psi _{1}(\alpha ,\beta ;\kappa ) &=&\frac{1}{\sqrt{2}}\left[ \left|
22\right\rangle -\left| 2-2\right\rangle \right] \sim {\rm cn}\,\alpha {\rm 
dn}\,\alpha {\rm sn}(\beta |\kappa ^{\prime }){\rm cn}(\beta |\kappa
^{\prime })  \nonumber \\
\Psi _{2}(\alpha ,\beta ;\kappa ) &=&\frac{1}{\sqrt{2}}\left[ \left|
21\right\rangle +\left| 2-1\right\rangle \right] \sim {\rm sn}\,\alpha {\rm 
cn}\,\alpha {\rm cn}(\beta |\kappa ^{\prime }){\rm dn}(\beta |\kappa
^{\prime })  \nonumber \\
\Psi _{3}(\alpha ,\beta ;\kappa ) &=&\frac{1}{\sqrt{2}}\left[ \left|
21\right\rangle -\left| 2-1\right\rangle \right] \sim {\rm sn}\,\alpha {\rm 
dn}\,\alpha {\rm sn}(\beta |\kappa ^{\prime }){\rm dn}(\beta |\kappa
^{\prime }) \\
\Psi _{4}(\alpha ,\beta ;\kappa ) &\sim &\left| 22\right\rangle +\left|
2-2\right\rangle +\sqrt{\frac{2}{3}}\frac{2f_{+}(\kappa )-1-\kappa ^{2}}{
1-\kappa ^{2}}\left| 20\right\rangle   \nonumber \\
&\sim &(1-f_{+}(\kappa ){\rm sn}\,^{2}\alpha )(1-f_{-}(\kappa ^{\prime })
{\rm sn}\,^{2}(\beta |\kappa ^{\prime }))  \nonumber \\
\Psi _{5}(\alpha ,\beta ;\kappa ) &\sim &\left| 22\right\rangle +\left|
2-2\right\rangle +\sqrt{\frac{2}{3}}\frac{2f_{-}(\kappa )-1-\kappa ^{2}}{
1-\kappa ^{2}}\left| 20\right\rangle   \nonumber \\
&\sim &(1-f_{-}(\kappa ){\rm sn}\,^{2}\alpha )(1-f_{+}(\kappa ^{\prime })
{\rm sn}\,^{2}(\beta |\kappa ^{\prime }))  \nonumber
\end{eqnarray}
with eigenvalues 
\begin{eqnarray}
E_{1} &=&1+\kappa ^{2},\qquad E_{2}=4+\kappa ^{2},\qquad E_{3}=1+4\kappa
^{2},\qquad  \\
E_{4} &=&2f_{+}(\kappa ),\qquad E_{5}=\ 2f_{-}(\kappa ),\qquad 
\end{eqnarray}
where $f_{\pm }(\kappa )=\left( 1+\kappa ^{2}\pm \sqrt{1-\kappa
    ^{2}\kappa ^{\prime 2}}\right) $. 
In general there will
be a relation between the eigenstates $\left| \ell
    m\right\rangle $ of the $2\ell+1$ band edges and
products of Lam\'e polynomials. The band edges for $\ell =1,2$ are shown in
Fig. 3 as a function of $\kappa ^{2}$. The bands are indicated by the shaded
regions. The dashed line indicates the height of the potential. One can see
that the bands merge at $\kappa =0$, and pass to the bound and scattering
states of the P\"oschl-Teller potential for $\kappa =1$.

\section{su(1,1) Realizations of the Lam\'e Equation}

\subsection{Elliptic Parametrization of su(1,1)}

In order to discuss the eigenstates in the band, and develop the dispersion
relation, we require more general representations. In the spirit of recent
work on the Scarf potential where it was shown that one can use a $su(1,1)$
dynamical symmetry to analytically solve for the dispersion relation\cite
{prllk}, states and transfer matrix, we consider transforming our generators
to $su(1,1)$. Consider the transformation of our previous coordinates
associated with $x\rightarrow -ix,\,y\rightarrow -iy:$ 
\begin{equation}
x=-i\kappa \,{\rm sn}\,\alpha \,{\rm sn}\,\beta ,\,y=\frac{\kappa }{\kappa
^{\prime }}\,{\rm cn}\,\alpha \,{\rm cn}\,\beta ,\,z=\frac{1}{\kappa
^{\prime }}\,{\rm dn}\,\alpha \,{\rm dn}\,\beta ,
\end{equation}
where $0\leq \alpha <4K,\,0\leq \beta <iK^{^{\prime }}$\cite{patera}.
This now parametrizes the hyperbolic surface $z^{2}-x^{2}-y^{2}=1$
illustrated in Fig. 4(a) for $\kappa ^{2}=1/2$. Contours are shown for
selected values of $\alpha $ and $\beta $ in Fig. 4(b). The generators of $
su(1,1)$ algebra are given in Table 2 as realization (I), and satisfy the
commutation relations $\left[ L_{x},L_{y}\right] =-L_{z}$, $\left[
L_{y},L_{z}\right] =L_{x}$, and $\left[ L_{z},L_{x}\right]
=L_{y}.$ The
Casimir invariant now has the form 
\begin{equation}
C_{2}=L_{z}^{2}-L_{x}^{2}-L_{y}^{2}=\frac{-1}{\kappa ^{2}({\rm sn}
^{2}\,\alpha -{\rm sn}^{2}\,\beta )}\left( \frac{\partial ^{2}}{\partial
\alpha ^{2}}-\frac{\partial ^{2}}{\partial \beta ^{2}}\right) 
\end{equation}
As in the $su(2)$ case, there are three forms of bilinear Hamiltonians which
lead to Schr\"odinger equations in $\alpha $ and $\beta $. These are given in
the top right of Table 2. If we choose \ $H_{3}=L_{z}^{2}-\kappa
^{2}L_{x}^{2}={\cal E}$, together with the Casimir invariant $C_{2}=-\ell
(\ell +1)$, we obtain the decoupled Lam\'e equations:
\begin{eqnarray}
\ \left[ -\frac{d^{2}}{d\alpha ^{2}}+\kappa ^{2}\ell (\ell +1){\rm sn}
^{2}\alpha \right] \psi (\alpha ) &=&-{\cal E}\psi (\alpha ),  \label{sua} \\
\ \left[ -\frac{d^{2}}{d\beta ^{2}}+\kappa ^{2}\ell (\ell +1){\rm sn}
^{2}\beta \right] \phi (\beta ) &=&-{\cal E}\phi (\beta ).  \label{sub}
\end{eqnarray}
Again similar results are obtained if we use $H_{1}$ or $H_{2}$, the only
difference arising in the definition of the eigenvalue. Note that the range
of $\beta $ is along the imaginary axis. This realization is
problematic in the  $\kappa =0$ and $\kappa =1$ limits since the
generators become singular. Consequently we consider a slightly different realization of
$su(1,1)$ below.

\subsection{Another su(1,1) Realization}

While we do not have a coordinate system which avoids the singularities at $
\kappa =1$, it is possible to at least allow a study of the scattering
states when $\kappa =0.$ To do this we make the transformations (i) $\beta
\rightarrow \beta +K+iK^{^{\prime }}$ followed by (ii) $\beta \rightarrow
-i\beta $. As a result: 
\begin{eqnarray}
{\rm sn}(\beta |\kappa ) &\rightarrow &\frac{1}{\kappa }{\rm dn}(\beta
|\kappa ^{\prime }),  \nonumber \\
{\rm cn}(\beta |\kappa ) &\rightarrow &\frac{\kappa ^{\prime }}{i\kappa }
{\rm cn}(\beta |\kappa ^{\prime }),  \nonumber \\
{\rm dn}(\beta |\kappa ) &\rightarrow &\kappa ^{\prime }{\rm sn}(\beta
|\kappa ^{\prime }).
\end{eqnarray}
This results in coordinates\cite{patera} 
\begin{equation}
x=-i\,{\rm sn}\,\alpha \,{\rm dn}(\beta |\kappa ^{\prime }),\,y=-i\,{\rm cn}
\,\alpha \,{\rm cn}(\beta |\kappa ^{\prime }),\,z={\rm dn}\,\alpha \,{\rm sn}
(\beta |\kappa ^{\prime }),
\end{equation}
where $0\leq \alpha <4K,\,-iK\leq \beta <-iK+K^{^{\prime }}$, also
satisfying $z^{2}-x^{2}-y^{2}=1$. Typical contours for this parametrization
are shown in Fig. 4(c). The generators are given in Table 2 as realization
(II) and satisfy $\left[ L_{x},L_{y}\right] =-L_{z}$, $\left[ L_{y},L_{z}
\right] =L_{x}$, and $\left[ L_{z},L_{x}\right] =L_{y}$ with Casimir
invariant 
\begin{equation}
C=L_{z}^{2}-L_{x}^{2}-L_{y}^{2}=\ \frac{1}{\kappa ^{2}{\rm cn}^{2}\alpha
+\kappa ^{\prime 2}{\rm cn}^{2}(\beta |\kappa ^{\prime })}\left( \frac{
\partial ^{2}}{\partial \alpha ^{2}}+\frac{\partial ^{2}}{\partial \beta ^{2}
}\right)
\end{equation}
\qquad The Hamiltonians which result in seperable Hamiltonians are given in
the bottom right of Table 2. Using $H_{1}=L_{x}^{2}+\kappa ^{\prime
2}L_{y}^{2}={\cal E}\ $and $C_{2}=-\ell (\ell +1)$ leads to the two Lam\'{e}
equations: 
\begin{eqnarray}
\ \left[ -\frac{d^{2}}{d\alpha ^{2}}+\kappa ^{2}\ell (\ell +1){\rm sn}
^{2}\alpha \right] \psi (\alpha ) &=&{\cal E}\psi (\alpha ),  \nonumber \\
\ \left[ -\frac{d^{2}}{d\beta ^{2}}+\kappa ^{\prime 2}\ell (\ell +1){\rm sn}
^{2}(\beta |\kappa ^{\prime })\right] \phi (\beta ) &=&(\ell (\ell +1)-{\cal 
E})\phi (\beta ).
\end{eqnarray}
It should be kept in mind that $\beta $ is still complex valued. Never the
less this realization allows an analysis of the $\kappa =0$ dynamical
symmetry.

\subsection{$\kappa =0$ Limit}

\bigskip

This realization is not singular in the $\kappa =0$ limit (although it is in
the $\kappa =1$ case). Taking $\kappa =0$, and shifting $\beta $ to be on
the real axis by $\beta =\theta -i\pi /2$, we find 
\begin{eqnarray}
L_{\pm } &=&\pm e^{\pm i\alpha }\left( \mp \sinh \theta \frac{\partial }{
\partial \theta }+i\cosh \theta \frac{\partial }{\partial \alpha }\right) 
\nonumber \\
L_{z} &=&-\ \frac{\partial }{\partial \alpha }.
\end{eqnarray}
To recover the usual $su(1,1)$ commutation relations $[I_{z},I_{\pm }]=\pm
I_{\pm }$, $[I_{+},I_{-}]=-2I_{z}$, we then identify $I_{\pm }=\pm L_{\pm }$
and $I_{z}=iL_{z}$.  It is convient to perform the
transformations: $\theta \rightarrow \theta -i\pi /2$, $\tanh \theta
\rightarrow \cos \theta $, followed by a similarity transformation $f(\theta
)=\sqrt{\sin \theta }$, and $\theta \rightarrow i\theta $. Then we have the
form: 
\begin{eqnarray}
\ -\frac{d^{2}}{d\alpha ^{2}}\ \psi (\alpha ) &=&m^{2}\psi (\alpha ), 
\nonumber \\
\ \left[ -\frac{d^{2}}{d\theta ^{2}}+\frac{m^{2}-1/4}{\sinh ^{2}\theta }
\right] \phi (\theta ) &=&-(\ell +\frac{1}{2})^{2}\phi (\theta ).
\end{eqnarray}
The principal series $\ell =-1/2+i\rho $ of the projective representations
of $su(1,1)$ now describe these scattering states, and the eigenfunctions
are of the form: 
\begin{equation}
\Psi _{\ell }^{m}\sim \sqrt{i\sinh \theta }P_{\ell }^{m}(\cosh \theta
)e^{\pm im\alpha }\quad \qquad m\in \Re ,\ \ell =-1/2+i\rho ,\ \rho >0.
\end{equation}

\section{Band Structure of the Lam\'{e} Hamiltonian}

We now focus on the properties of the Lam\'e equation in the form of Eq. (1).
As we are interested in band structure, \ the eigenstates must satisfy
Bloch's theorem. For a potential which is periodic with period $a$, $
V(x+a)=V(x)$, the wavefunctions must be of the form 
\begin{equation}
\Psi _{k}(x)=u_{k}(x)\exp [-ikx],
\end{equation}
where $k$ is the wavenumber and $u_{k}(x)$ has the periodicity of the
lattice: $u_{k}(x+a)=u_{k}(x)$. As the eigenstates $\Psi _{k}(x)$ are not
periodic, the doubly periodic solutions of the
Lam\'e equation do not play a role for energies in the band. Rather, we look
to the more general class of solutions expressed in terms of Jacobi theta
functions\cite{whitwat}. Starting with the Hamiltonian 
\begin{equation}
H\psi =\left[ -\frac{d^{2}}{dx^{2}}+\kappa ^{2}\ell (\ell +1){\rm sn}
^{2}(x|\kappa )\right] \psi (x)={\cal E}\psi (x),  \label{Lameham}
\end{equation}
the solutions for positive integer $\ell $ are given parametrically by

\begin{equation}
\psi (x)=\prod_{n=1}^{\ell }\left[ \frac{{\cal H}(x+\alpha _{n})}{\theta (x)}
e^{-xZ(\alpha _{n})}\right]
\end{equation}
where ${\cal H}$ and $\theta$ are theta functions, and
$\alpha _{1},\alpha _{2},...\alpha _{\ell }$ are constants
determined by the constraints: 
\begin{eqnarray}
{\cal E} &=&\sum_{n=1}^{\ell }\frac{1}{{\rm sn}^{2}\alpha _{n}}-\left[ \sum_{n=1}^{\ell }
{\rm cn}\alpha _{n}{\rm dn}\alpha _{n}/{\rm sn}\alpha _{n}\right] ^{2}
\label{eq:const} \\
0 &=&\sum_{p=1}^{\ell }\frac{{\rm sn}\alpha _{p}{\rm cn}\alpha _{p}{\rm dn}
\alpha _{p}+{\rm sn}\alpha _{n}{\rm cn}\alpha _{n}{\rm dn}\alpha _{n}}{{\rm 
sn}^{2}\alpha _{p}-{\rm sn}^{2}\alpha _{n}}\quad (p\neq n)
\end{eqnarray}
If this solution is not doubly periodic, a second solution is

\begin{equation}
\psi ^{^{\prime }}(x)=\prod_{n=1}^{\ell }\left[ \frac{{\cal H}(x-\alpha _{n})}{
\theta (x)}e^{xZ(\alpha _{n})}\right].
\end{equation}
We can then identify the dispersion relation by putting these
wavefunctions in Bloch form and by using the periodicity of the
theta functions to extract $u_k(x)$. We find

\begin{equation}
k({\cal E})=-i\sum_{n=1}^{\ell }Z(\alpha _{n}|\kappa ^{2})+\frac{\ell \pi }{2K}.
\label{eq:disp}
\end{equation}
We will start with the case of $\ell =1$, where simple anlytic
results can be obtained.  We then derive the transfer matrix for
the general case of integer $\ell .$

\subsection{$ \ell =1$ Results}

Since there is only one parameter $\alpha ,$ the constraint equation is
simply

\begin{equation}
{\rm dn}^{2}\alpha ={\cal E}-\kappa ^{2}.
\end{equation}
The condition that the dispersion relation is real,
Re$Z(\alpha|\kappa^2)=0$, results in two energy bands given by 
\begin{equation}
\kappa ^{2}\leq {\cal E}\leq 1,\quad 1+\kappa ^{2}\leq {\cal E}. 
\end{equation}
These are shown in Fig. 3 (top). In the lower band, $\alpha $ has the form $
\alpha =K+i\eta $, where $\eta $ ranges from $K^{\prime }$ at ${\cal E}
=\kappa ^{2}$, to $0$ at ${\cal E}=1.$ In the upper band, $\alpha =i\eta $,
where $\eta $ ranges from $0$ at ${\cal E}=1+\kappa ^{2}$, to $K^{\prime }$
as ${\cal E}\rightarrow \infty $. This path traced out by the
parameter $\alpha$ as a function energy ${\cal E}$ is shown schematically in Fig. 5. The upper
and lower sides correspond to band gaps while the right and left
edges are the energy bands.
Using the specific forms of $\alpha $ for each band, the dispersion relation
becomes 
\begin{equation}
k({\cal E})=\left\{ 
\begin{array}{ll}
-Z(\eta |\kappa ^{\prime 2})+\frac{\pi }{2K}(1-\frac{\eta }{K^{\prime }})+
\sqrt{\frac{({\cal E}-\kappa ^{2})(1-{\cal E)}}{1+\kappa ^{2}-{\cal E}}} & 
\kappa ^{2}\leq {\cal E}\leq 1, \\ 
-Z(\eta |\kappa ^{\prime 2})+\frac{\pi }{2K}(1-\frac{\eta }{K^{\prime }})+
\sqrt{\frac{({\cal E}-\kappa ^{2}-1)({\cal E}-\kappa ^{2})}{{\cal E}-1}} & 
1+\kappa ^{2}\leq {\cal E}
\end{array}
\right.
\end{equation}
This is plotted in Fig. 6 (a) for the case of $\ell=1$. The
momentum $k$ is plotted up to the edge of the Brillouin zone,
which is $k=\pi/2K$. In the figure we use $\kappa^2=1/2$, so
that the band edges are ${\cal E}=1/2$, 1 and 3/2. (The
analogous behavior for $\ell=2$ is indicated in Fig. 6(b).)
The solution of the Lam\'e equation in Bloch form is now 
\begin{equation}
\psi _{k}(x)=\left\{ 
\begin{array}{ll}
\left[ \frac{H_{1}(x+i\eta )}{\Theta (x)}\exp (i\pi x/2K)\right] \exp (-ikx)
& \kappa ^{2}\leq {\cal E}\leq 1, \\ 
\left[ \frac{H(x+i\eta )}{\Theta (x)}\exp (i\pi x/2K)\right] \exp (-ikx) & 
1+\kappa ^{2}\leq {\cal E}
\end{array}
\right. .
\end{equation}
The component of the wavefunction in square brackets can be checked to be
periodic with the periodicity of the direct lattice: $x\rightarrow x+2K.$

The dispersion relation displays the desired limits. One can see that $k(
{\cal E})\rightarrow 0$ as ${\cal E}\rightarrow \kappa ^{2}$, and $k({\cal E}
)\rightarrow \pi /2K$ as ${\cal E}\rightarrow 1.$ Further, as the modulus of
the elliptic function vanishes, $\kappa ^{2}\rightarrow 0$, the Hamiltonian
becomes that of a free system, and we find $k({\cal E})\rightarrow \sqrt{
{\cal E}}$ as desired. In the P\"oschl-Teller limit, $\kappa ^{2}\rightarrow 1$
, the band vanishes, and we find $k({\cal E})\rightarrow 0$. Similar
results hold for the upper band as well.

From the dispersion relation we can compute the group velocity and effective
mass. To do so, we use the relation between the zeta function and the
elliptic integrals of the first and second kind

\begin{equation}
Z(\alpha )=E(\alpha )-\frac{E(\kappa ^{2})}{K}\alpha
\end{equation}
where $E(\kappa ^{2})=\frac{\pi }{2}F(-1/2,1/2;1;\kappa ^{2})$ and $K=\frac{
\pi }{2}F(1/2,1/2;1;\kappa ^{2})$ are the complete elliptic integrals and $
E(\alpha )$ is incomplete. Then, the group velocity is given by:

\begin{equation}
\frac{1}{\nu }=\frac{dk({\cal E})}{d{\cal E}}
\end{equation}
or

\begin{equation}
\nu ({\cal E})=\frac{\ 2\sqrt{(1-{\cal E})({\cal E}-\kappa ^{2})(1+\kappa
^{2}-{\cal E})}}{\kappa ^{2}+\frac{E(\kappa ^{2})}{K}-{\cal E}}.
\end{equation}
We plot $\nu ^{2}$ as a function of energy in Fig. 7 for several values of $
\kappa .$ As $\kappa$ approaches zero, the energy gaps vanish,
and the group velocity approaches the free particle limit $E=M\nu^2/2=\nu^2/4$
(dot-dashed line). As $\kappa$ approaches unity, the lower band
vanishes becoming a  bound state, and the group velocity is only
non-vanishing for the continuum states of the P\"oschl-Teller
potential with ${\cal E}\ge 2$. 

The effective mass $M^{\ast }$ is determined by

\begin{eqnarray}
\frac{1}{M^{\ast }} &=&\frac{d\nu }{dk}  \nonumber \\
&=&-2\frac{({\cal E}-\kappa ^{2})(1+\kappa ^{2}-{\cal E})+({\cal E}
-1)(1+\kappa ^{2}-{\cal E})-({\cal E}-1)({\cal E}-\kappa ^{2})}{({\cal E}
-\kappa ^{2}-\frac{E(\kappa ^{2})}{K})^{2}}  \nonumber \\
&&+4\frac{({\cal E}-1)({\cal E}-\kappa ^{2})(1+\kappa ^{2}-{\cal E})}{({\cal 
E}-\kappa ^{2}-\frac{E(\kappa ^{2})}{K})^{3}}
\end{eqnarray}
We plot $1/M^*$  in Fig. 8 for selected values of $\kappa .$ One can see that
as $\kappa ^{2}\rightarrow 0$, the gaps vanish and $M^{\ast }\rightarrow M=1/2$ as expected for
the free particle.

\subsection{Transfer Matrix for the Lam\'e Hamiltonian}

The general form of the transfer matrix is computed using the
definitions in the Appendix. Using the wavefunctions and Eq. (A1), we have 
\begin{equation}
r=-\sum_{n=1}^{\ell }[Z(\alpha _{n})-\frac{{\rm sn}\,\alpha _{n}\,{\rm dn}
\,\alpha _{n}}{{\rm cn}\,\alpha _{n}}]-i\frac{\ell \pi }{2k}.
\end{equation}
Consequently, we see that 
\begin{equation}
r+ik({\cal E})=\sum_{n=1}^{\ell }\frac{{\rm sn}\,\alpha _{n}\,{\rm dn}
\,\alpha _{n}}{{\rm cn}\,\alpha _{n}}.
\end{equation}
The transfer matrix then has the form 
\begin{equation}
T\,=\,\left( 
\begin{array}{cc}
\cos 2k({\cal E})K & i\left( \sum_{n=1}^{\ell }\frac{{\rm sn}\,\alpha _{n}\,
{\rm dn}\,\alpha _{n}}{{\rm cn}\,\alpha _{n}}\right) ^{-1}\sin 2k({\cal E})K
\\ 
i(\sum_{n=1}^{\ell }\frac{{\rm sn}\,\alpha _{n}\,{\rm dn}\,\alpha _{n}}{{\rm 
cn}\,\alpha _{n}})\sin 2k({\cal E})K & \cos 2k({\cal E})K
\end{array}
\right) .
\end{equation}
(Note that this is in the form of Eq. (A.3) rather than (A.5).)
While this expression requires knowledge of the parameters $\alpha _{n}$,
one can obtain various limits of this for the free particle and
P\"oschl-Teller potentials.

\subsection{The $\kappa =0$ Free Particle Limit}

We start first with the $\ell =1$ case. Taking $\kappa =0$ in our transfer
matrix, we obtain for the upper and lower bands
\begin{equation}
T\,=\,\left( 
\begin{array}{cc}
\cosh 2k({\cal E})K & i\frac{{\rm cn}\,\alpha }{{\rm sn}\,\alpha \,{\rm dn}
\,\alpha }\ \sinh 2k({\cal E})K \\ 
i\frac{{\rm sn}\,\alpha \,{\rm dn}\,\alpha }{{\rm cn}\,\alpha }\sinh 2k(
{\cal E})K & \cosh 2k({\cal E})K\ 
\end{array}
\right) .
\end{equation}
$\ $According to Eq. (\ref{eq:const}), 
\begin{equation}
\frac{{\rm sn}\,\alpha \,{\rm dn}\,\alpha }{{\rm cn}\,\alpha }=\left[ \frac{
(1+\kappa ^{2}-{\cal E})({\cal E}-\kappa ^{2})}{({\cal E}-1)}\right] ^{1/2}.
\label{eq:ratio}
\end{equation}
Taking $\kappa =0$, we have $r\rightarrow 0$, so that 
\begin{equation}
k({\cal E})=-i\frac{{\rm sn}\,\alpha \,{\rm dn}\,\alpha }{{\rm cn}\,\alpha }=
\sqrt{{\cal E}}.
\end{equation}
The transfer matrix  becomes that for a free particle, given by: 
\begin{equation}
T\,=\,\left( 
\begin{array}{cc}
\cos \pi \sqrt{{\cal E}} & {\cal E}^{-1/2}\sin \pi \sqrt{{\cal E}} \\ 
-{\cal E}^{1/2}\sin \pi \sqrt{{\cal E}}\  & \ \cos \pi \sqrt{{\cal E}}
\end{array}
\right) .  \label{eq:freetrans}
\end{equation}
 For the general case of integer $\ell $, we start with
the constraint equations (\ref{eq:const}) and take the limit $\kappa
\rightarrow 0.$  We first must show that 
\begin{equation}
r+ik({\cal E})=\sum_{n=1}^{\ell }\tan \alpha _{n}\rightarrow i\sqrt{{\cal E}}
.  \label{eq:sumtan}
\end{equation}
Then, from the definition of $r$ in the appendix, it must vanish for free
motion, so that $r\rightarrow 0$ implies $k({\cal E})\rightarrow \sqrt{{\cal 
E}}$. Then we recover the free particle transfer matrix. While we can show
that the sum in Eq. (\ref{eq:sumtan}) tends to $i\sqrt{{\cal E}}$ for small $
\ell $ on a case by case basis, we do not yet have a general proof. However,
since the transfer matrix must be that of a free particle in this limit, it
is clear that Eq. (\ref{eq:sumtan}) must hold, and we can use this instead
to provide an additional relation among the parameters $\alpha _{n}$.

\subsection{The $\kappa =1$ P\"oschl-Teller Limit}

\bigskip

In the limit $\kappa =1$, the Hamiltonian $H$ becomes the P\"oschl-Teller
Hamiltonian, and our transfer matrix should reduce to that case. We start
first with $\ell =1$ and examine the upper band.(The lower band becomes
degenerate at $\kappa =1$). In this limit, $K\rightarrow \infty $ and $
K^{\prime }\rightarrow \pi /2$. For the upper band, 
\begin{equation}
k({\cal E})\rightarrow \sqrt{{\cal E}-2}+\frac{\pi +2i\alpha }{2K},
\end{equation}
From Eq. (\ref{eq:ratio}), we have in the $\kappa =1$ limit: 
\begin{equation}
\frac{{\rm sn}\,\alpha \,{\rm dn}\,\alpha }{{\rm cn}\,\alpha }=\tanh \alpha
=i\sqrt{{\cal E}-2}.  \label{leq:ratioa}
\end{equation}
The asymptotic form of the transfer matrix (as $K\rightarrow \infty $)
becomes: 
\begin{equation}
T\,=\,\left( 
\begin{array}{cc}
\cos (2K\sqrt{{\cal E}-2}+2i\alpha ) & ({\cal E}-2)^{-1/2}\sin (2K\sqrt{
{\cal E}-2}+2i\alpha ) \\ 
-\ ({\cal E}-2)^{1/2}\sin (2K\sqrt{{\cal E}-2}+2i\alpha ) & \cos (2K\sqrt{
{\cal E}-2}+2i\alpha )\ 
\end{array}
\right) .
\end{equation}
We must now change this form of the transfer matrix, defined for periodic
potentials $T$, to the form used in the P\"oschl-Teller case which is not
periodic, ${\cal T}\,$(see Appendix). Using $k({\cal E})=\sqrt{{\cal E}-2}$
, we find: 
\begin{equation}
{\cal T}\,_{\kappa =1}=\,\left( 
\begin{array}{cc}
\frac{\Gamma (ik)\Gamma (1+ik)}{\Gamma (2+ik)\Gamma (ik-1)} & \ 0 \\ 
0 & \ \frac{\Gamma (-ik)\Gamma (1-ik)}{\Gamma (2-ik)\Gamma (-ik-1)}
\end{array}
\right) .
\end{equation}

For the case of arbitrary $\ell $, we have the more general relations: 
\begin{equation}
\sum_{n=1}^{\ell }\frac{\,{\rm sn}\alpha _{n}{\rm cn}\alpha _{n}}{{\rm dn}
\alpha _{n}}\rightarrow \sum_{n=1}^{\ell }\tanh \alpha _{n}=i\sqrt{{\cal E}
-\ell (\ell +1)}.
\end{equation}
The dispersion relation becomes 
\begin{equation}
k({\cal E})\rightarrow -i\sum_{n=1}^{\ell }(\tanh \alpha _{n}-\frac{\alpha
_{n}}{K})+\ell \frac{\pi }{2K}=\sqrt{{\cal E}-\ell (\ell +1)}+\ell \frac{\pi 
}{2K}+i\frac{1}{K}\sum_{n=1}^{\ell }\alpha _{n}.
\end{equation}
To compute the form of the transfer matrix for the P\"oschl-Teller potential,
we use the notation

\begin{equation}
{\cal T}\,=\,\left( 
\begin{array}{cc}
F & G\  \\ 
G^{\ast } & F^{\ast }
\end{array}
\right) .
\end{equation}
where, from the appendix, we have: 
\begin{equation}
F=(-)^{\ell }\exp \left[ -2\sum_{n=1}^{\ell }\alpha _{n}\right]
=\prod_{n=1}^{\ell }\frac{\tanh \alpha _{n}-1}{\tanh \alpha _{n}+1}.
\end{equation}
This can be further simplified using the limiting form of the constraint
equations (\ref{eq:const}): 
\begin{eqnarray}
{\cal E} &=&-\left[ \sum_{n=1}^{\ell }\frac{1-\tanh ^{2}\alpha _{n}}{\tanh
\alpha _{n}}\right] ^{2}+\sum_{n=1}^{\ell }\coth ^{2}\alpha _{n}\ , 
\nonumber \\
0 &=&\sum_{p=1(p\neq n)}^{\ell }\frac{(1-\tanh ^{2}\alpha _{n})\tanh \alpha
_{n}+(1-\tanh ^{2}\alpha _{p})\tanh \alpha _{p}}{\tanh ^{2}\alpha _{p}-\tanh
^{2}\alpha _{n}}\ .
\end{eqnarray}
For small $\ell $ ($\ell =1,2$) we can solve these equations to show that 
\begin{equation}
F=\frac{\Gamma (ik)\Gamma (1+ik)}{\Gamma (1+ik+\ell )\Gamma (ik-\ell )}
,\qquad G=0.
\end{equation}
This reduces to the correct form of the transfer matrix. For arbitrary $\ell 
$, we have not been able to solve these equations, but the result must still
hold, since it is governed by the form of the Hamiltonian. Again, we can use
these relations to provide additional relations between the parameters $
\alpha _{n}$.

\section{Conclusions}

We have presented a group theoretical analysis of the Lam\'e equation, which
is an example of a SGA band structure problem for $su(2)$ and $su(1,1)$. We
have computed the dispersion relation and transfer matrix, and discussed the
limiting dynamical symmetry limits of these results which correspond to the
P\"oschl-Teller and free particle Hamiltonians. Because the general
Hamiltonian is not of the dynamical symmetry the spectrum cannot be obtained
in closed form. Never the less, a diagonalization of Hamiltonians which are
bilinear in the angular momentum generators will provide the general
solution. There are still many open questions associtated with the group
theoretical treatment of the Lam\'e equation. It would be nice to develop a $
su(1,1)$ parametrization which is non-singular for all values of $\kappa .$
 In addition, the
Scarf and Mathieu equation limits would be interesting to realize more expliticly in
the transfer matrix and dispersion relations. Finally, the diagonalization
of the algebraic Hamiltonians in the continuum $su(1,1)$ bases would be
interesting to study. It is clear that the result of the diagonalization
must yield the same transcendental equations for the parameters $\alpha _{n}$
, but their origin would be different.

\vspace{2cm}

\section*{\bf Appendix: Form of the transfer matrix for periodic potentials}
\setcounter{equation}{1}
\renewcommand{\theequation}{A.\arabic{equation}}

For symmetric periodic potentials with period $\tau $, we can derive a
general formular of the transfer matrix. Suppose for a specific energy E, we
have two bloch solutions $u_{+}(x)e^{ikx}$, $u_{-}(x)e^{-ikx}$, where $x$ is
the coordinate, $k=k({\cal E})$ is the dispersion relation. Since the
potential is symmetric, we can define\cite{James} 
\begin{equation}
r=\frac{u_{+}^{^{\prime }}(\tau /2)}{u_{+}(\tau /2)}=-\frac{u_{-}^{^{\prime
}}(\tau /2)}{u_{-}(\tau /2)}.
\end{equation}
Then the transfer matrix is 
\begin{equation}
T\,=\,\left( 
\begin{array}{cc}
\cos k\tau  & \frac{i}{ik+r}\sin k\tau  \\ 
i(ik+r)\sin k\tau  & \cos k\tau 
\end{array}
\right) .
\end{equation}
The forms of transfer matrices used for periodic and non-periodic potentials
is different. In the discussion of the Lam\'e equation, the $\kappa
\rightarrow 1$ limit takes a periodic potential to a non-periodic one, so
that we require the transformations that take us from one standard form to
the other. If we express the transfer matrix $T$ for the periodic potential
as 
\begin{equation}
T=\left( 
\begin{array}{cc}
{\rm Re}(F\exp (ik\tau )+G) & \frac{1}{k}{\rm Im}(F\exp (ik\tau )+G) \\ 
k{\rm Im}(F\exp (ik\tau )-G)\  & {\rm Re}(F\exp (ik\tau )-G)\ 
\end{array}
\right) ,
\end{equation}
then that of the non-periodic limit will have the form 
\begin{equation}
{\cal T}=\left( 
\begin{array}{cc}
F & G \\ 
G^{\ast }\  & F^{\ast }
\end{array}
\right) .
\end{equation}
In this notation, $k=k({\cal E})$ is the dispersion relation, and $\tau $ is
the period of the periodic potential which tends to $\infty .$

\newpage

\begin{center}
Figure Captions
\end{center}

\begin{enumerate}
\item[Figure 1.] Various forms of the Lam\'e potential which are real valued.
We plot $V(x)$ for modulus $\kappa ^{2}=1/2$ given by {\it (i)} ${\rm sn}
^{2}(x|\kappa )$ (solid), {\it (ii)}$\ $ ${\rm sn}^{2}(x+iK^{\prime }|\kappa
)$ (dots), {\it (iii)}$\ $ ${\rm sn}^{2}(x+K+iK^{\prime }|\kappa )$
(dashes), {\it (iv)}$\ {\rm sn}^{2}(ix|\kappa )$ (dot-dashes). For other
values of $x$, the potential is periodic, but generally complex.

\item[Figure 2.] Coordinate systems for $su(2)$. (a) Parametrization of the
sphere. (b) Contours of constant $\alpha$ and $\beta$. (c) Same as (b) but
for the second parametrization.

\item[Figure 3.] Band edges from the $su(2)$ realization of the Lam\'e
equation as a function of $\kappa ^{2}$ for $\ell =1$ (top) and $\ell =2$
(bottom). The eigenvalues are indicated. The bands are given by the shaded
regions. The dashed line indicates the value of the energy which is at the
maximum of the potential $V(x)=\kappa ^{2}\ell (\ell +1){\rm sn}^{2}x$

\item[Figure 4.] Coordinate systems for $su(1,1)$. (a) Parametrization of
the hyperboloid. (b) Contours of constant $\alpha$ and $\beta$. (c) Same as
(b) but for the second parametrization.

\item[Figure 5.] Evolution of the parameter $\alpha $ for the
  $\ell =1$ Lam\'{e} equation. The lower energy band has
  parameter $\alpha =K+i\eta $ where 
$\eta $ ranges from $K^{\prime }$ at the lower end to $0$ at the upper end.
The valence band starts with $\alpha =0$ and grows to $\alpha =iK^{\prime }$
as the energy $E\rightarrow \infty .$

\item[Figure 6.] (a) Dispersion relation ${\cal E}(k)$ for the $\ell=1$ Lam\'e
  equation with $\kappa^2=1/2$. The momentum is plotted in units
  of $\pi/2K$. (b) Analogous behavior for the $\ell=2$ case.

\item[Figure 7.] Group velocity for the $\ell =1$ Lam\'{e} equation for
several values of the modulus $\kappa $. We show $v^{2}$ as a function of
energy ${\cal E}$. When $\kappa =0$, the Hamiltonian is that of a free
particle so ${\cal E}\propto v^{2}$ (dot-dashes). As $\kappa $ increases to
unity, the lower energy band shrinks to a single bound state of multiplity 2
corresponding to the P\"oschl-Teller potential, and the group velocity is only
non-zero in the continuum, ${\cal E}>2$.

\item[Figure 8.] Behavior of $1/M^{\ast }$ as a function of energy ${\cal E}$
for $\ell =1$ Lam\'e equation, where $M^{\ast }$ is the effective mass. As $
\kappa \rightarrow 0$, the gap closes, and $M^{\ast }\rightarrow M=1/2$ for
this Hamiltonian.
\end{enumerate}
\newpage

\sideways

\begin{tabular}{ll|ll}
\multicolumn{4}{l}{Table 1. Realizations  of $SU(2)$ with
  $[L_i,L_j]=\epsilon_{ijk}L_k$ and corresponding
Hamiltonians which lead to the Lam\'e equation. }\\
\multicolumn{4}{l}{ }\\
\multicolumn{4}{l}{ }\\
&$SU(2)$ Realization & & Hamiltonians \\ \hline\hline
& & & \\
(I): && & \\
& & &\\
$L_{x}$& $ =i\left[\kappa ({\rm sn}^{2}\alpha 
-{\rm sn}^{2}\beta)\right]^{-1}$ & $H_1$ & $= L_{x}^{2}+\kappa
^{\prime 2}L_{y}^{2}$\\ 
& $\times
\left[ {\rm sn}\,\alpha \,{\rm cn}\,\beta \,
{\rm dn}\,\beta \partial_\alpha -\,{\rm cn}
\,\alpha \,{\rm dn}\,\alpha \,{\rm sn}\,\beta 
\partial_\beta \right] $ & &$ = 
 \left[\kappa^2({\rm sn}^{2}\,\alpha -{\rm sn}\,^{2}\beta) \right]^{-1}
\left( {\rm dn}^{2}\,\beta \partial_\alpha
^{2} -{\rm dn}^{2}\,\alpha \partial_\beta ^{2}\right)$\\
$L_{y}$ & $ =\left[\kappa \kappa ^{\prime }({\rm sn}^{2}\alpha -{\rm 
sn}^{2}\beta) \right]^{-1}$ &   $H_2$ & $ = \kappa^2 L_x^2 -
\kappa^{\prime 2} L_z^2$\\
& $\times \left[ {\rm cn}\,\alpha \,{\rm sn}\,\beta
\,{\rm dn}\,\beta \partial_\alpha -\,{\rm 
{\rm sn}}\,\alpha \,{\rm dn}\,\alpha \,{\rm cn}
\,\beta \partial_\beta \right]$ & & $ =  \left[{\rm
  sn}^{2}\,\beta -{\rm sn}\,^{2}\alpha\right]^{-1}
   \left( {\rm cn}^{2}\,\beta \partial_\alpha
  ^{2} -{\rm cn}^{2}\,\alpha \partial_\beta ^{2}\right)$\\
$L_{z}$ & $ = i\left[\kappa ^{\prime }({\rm sn}^{2}\alpha -{\rm sn}
^{2}\beta )\right]^{-1}$ & $H_3$ & $= L_{z}^{2}+\kappa ^{2}L_{y}^{2} $\\
& $\times \left[ {\rm dn}\,\alpha \,{\rm sn}\,\beta \,
{\rm cn}\,\beta \partial_\alpha -\,{\rm {\rm 
sn}}\,\alpha \,{\rm cn}\,\alpha \,{\rm dn}\,\beta\partial_\beta
\right] $ & & 
   $ = \left[{\rm sn}^{2}\,\beta -{\rm sn}\,^{2}\alpha \right]^{-1}
     \left( {\rm sn}^{2}\,\beta \partial_\alpha^{2} 
     -{\rm sn}^{2}\,\alpha \partial_\beta ^{2}\right)$ \\
 & &&\\ 
$C_2$ & $ =L_{x}^{2}+L_{y}^{2}+L_{z}^{2}=\left[\kappa ^{2}({\rm sn}^{2}\,\beta -{\rm sn}
\,^{2}\alpha )\right]^{-1}\left( \partial_\alpha ^{2}-\partial_\beta ^{2}\right)$ & &\\
    & &&\\\hline\hline
&&&\\
(II): &&& \\
& & &\\
$L_{x}$ & $ =\left[\kappa ^{2}{\rm cn}^{2}\alpha +\kappa ^{\prime 2}
{\rm cn}^{2}(\beta |\kappa ^{\prime })\right]^{-1} $ & $H_1$ & $
             =L_{x}^{2}+\kappa ^{\prime 2}L_{y}^{2}$\\
& $\times\left[ \kappa ^{\prime 2}\,{\rm sn}\,\alpha {\rm sn}
(\beta |\kappa ^{\prime }){\rm cn}(\beta |\kappa ^{\prime
  })\partial_\alpha 
 +\,{\rm cn}\,\alpha \,{\rm {\rm dn}
}\,\alpha \,{\rm dn}(\beta |\kappa ^{\prime })
\partial_\beta \right] $ & & $ =\left[\kappa ^{2}{\rm
cn}^{2}\alpha +\kappa ^{\prime 2}{\rm cn}^{2}(\beta
|\kappa ^{\prime })\right]^{-1}\left( \kappa ^{\prime 2}{\rm sn}^{2}\,(\beta
|\kappa ^{\prime })\partial_\alpha ^{2}+{\rm 
dn}\,^{2}\alpha \partial_\beta^{2}\right) $\\ 
$L_{y}$ & $=\left[\kappa ^{2}{\rm cn}^{2}\alpha +\kappa ^{\prime 2}
{\rm cn}^{2}(\beta |\kappa ^{\prime })\right]^{-1}$ & $H_2$ & $=\kappa
^{2}L_{x}^{2}-\kappa ^{\prime 2}L_{z}^{2}$\\
& $\times\left[ -{\rm cn}\,\alpha {\rm sn}(\beta |\kappa
^{\prime }){\rm dn}(\beta |\kappa ^{\prime })
\partial_\alpha +\,{\rm sn}\,\alpha \,{\rm dn}\,\alpha
\,{\rm cn}(\beta |\kappa ^{\prime })\partial
\beta \right]$ & & $=-\left[\kappa
^{2}{\rm cn}^{2}\alpha +\kappa ^{\prime 2}{\rm cn}
^{2}(\beta |\kappa ^{\prime })\right]^{-1}\left( \kappa ^{\prime 2}{\rm cn}
^{2}\,(\beta |\kappa ^{\prime })\partial_\alpha ^{2}
-\kappa ^{2}{\rm cn}\,^{2}\alpha\partial_\beta^{2}\right)$\\
$L_{z}$ & $=\left[\kappa ^{2}{\rm cn}^{2}\alpha +\kappa ^{\prime 2}
{\rm cn}^{2}(\beta |\kappa ^{\prime })\right]^{-1}$ & $H_3$ &
 $=L_{z}^{2}+\kappa ^{2}L_{y}^{2}$\\
&$\times\left[ -\,{\rm dn}\,\alpha \,{\rm cn}(\beta |\kappa
^{\prime }){\rm dn}(\beta |\kappa ^{\prime })
\partial_\alpha -\kappa ^{2}\,{\rm sn}\,\alpha \,{\rm {\rm cn}
}\,\alpha \,{\rm sn}(\beta |\kappa ^{\prime })
\partial_\beta \right]$ & & $=\left[\kappa
^{2}{\rm cn}^{2}\alpha +\kappa ^{\prime 2}{\rm cn}
^{2}(\beta |\kappa ^{\prime })\right]^{-1}\left({\rm dn}
^{2}\,(\beta |\kappa ^{\prime })\partial_\alpha ^{2}
+\kappa ^{2}{\rm sn}\,^{2}\alpha\partial_\beta^{2}\right)$\\

& & &\\
$C_2$ & $=L_{x}^{2}+L_{y}^{2}+L_{z}^{2}$ & &\\
& $=\left[\kappa ^{2}{\rm cn}
^{2}\alpha +\kappa ^{\prime 2}{\rm cn}^{2}(\beta |\kappa ^{\prime
})\right]^{-1}\left( \partial_\alpha ^{2}+{\rm \ }\partial_\beta ^{2}\right) $ & &\\
& & &\\
 \hline\hline
\end{tabular}
\endsideways

\newpage

\sideways

\begin{tabular}{ll|ll}
\multicolumn{4}{l}{Table 2. Realizations  of $SU(1,1)$ and corresponding
Hamiltonians which lead to the Lam\'e equation.}\\
\multicolumn{4}{l}{ }\\
\multicolumn{4}{l}{ }\\
&$SU(1,1)$ Realization & & Hamiltonians \\ \hline\hline
& & & \\
(I): && & \\
& & &\\
$L_{x}$ & $=i\left(\kappa \kappa ^{\prime }({\rm sn}^{2}\beta -{\rm
    sn}^{2}\alpha )\right)^{-1}$ & $H_1$ & $=\kappa
    ^{2}L_{y}^{2}+\kappa ^{\prime 2}L_{z}^{2}$\\
 & $\times \left[ {\rm cn}\,\alpha \,{\rm sn}\,\beta
\,{\rm dn}\,\beta \partial_\alpha  -\,{\rm 
{\rm sn}}\,\alpha \,{\rm dn}\,\alpha \,{\rm cn}
\,\beta \partial_\beta  \right]$ & & $= \left[{\rm 
{\rm sn}}^{2}\,\alpha -{\rm sn}^{2}\,\beta \right]^{-1}\left( {\rm 
{\rm cn}}^{2}\beta \partial ^{2}_\alpha-{\rm 
{\rm cn}}^{2}\alpha \partial ^{2}_\beta\right)$\\
$L_y$ & $=\left(\kappa ({\rm sn}^{2}\beta -{\rm sn}^{2}\alpha
  )\right)^{-1}$ & $H_2$ & $=L_{y}^{2}+\kappa ^{\prime
  2}L_{x}^{2}$ \\
 & $\times \left[ {\rm sn}\,\alpha \,{\rm cn}\,\beta \,{\rm 
{\rm dn}}\,\beta\partial_\alpha -\,{\rm cn}
\,\alpha \,{\rm dn}\,\alpha \,{\rm sn}\,\beta 
\partial_\beta \right]$ & & $=\left[\kappa ^{2}({\rm 
{\rm sn}}^{2}\,\alpha -{\rm sn}^{2}\,\beta )\right]^{-1}\left( {\rm 
{\rm dn}}^{2}\beta \partial_\alpha ^{2}-\ {\rm 
{\rm dn}}\,^{2}\alpha \partial_\beta^{2}\right) $\\
$L_z$ & $=i\left[\kappa ^{\prime }({\rm sn}^{2}\beta -{\rm sn}
^{2}\alpha )\right]^{-1}$ & $H_3$ & $= L_z^2-\kappa^2 L_x^2$\\
 & $\times\left[ {\rm dn}\,\alpha \,{\rm sn}\,\beta \,
{\rm cn}\,\beta \partial_\alpha -\,{\rm {\rm 
sn}}\,\alpha \,{\rm cn}\,\alpha \,{\rm dn}\,\beta 
\partial_\beta \right]$ &  & $= -\left[{\rm sn}
^{2}\,\alpha -{\rm sn}^{2}\,\beta \right]^{-1}\left( {\rm \ {\rm sn}}
^{2}\beta \partial_\alpha^{2}-\ {\rm sn}
^{2}\alpha \partial_\beta^{2}\right)$\\
& & &\\
$C_2$ & $=L_{z}^{2}-L_{x}^{2}-L_{y}^{2}=-\left[\kappa ^{2}({\rm sn}
^{2}\,\alpha -{\rm sn}^{2}\,\beta )\right]^{-1}\left(
\partial_\alpha ^{2} - \partial_\beta
^{2}\right)$ & &\\
    & &&\\\hline\hline
&&&\\
(II): &&& \\
& & &\\
$L_{x}$ & $=i\left[\kappa ^{2}{\rm cn}^{2}\alpha +\kappa ^{\prime 2}
{\rm cn}^{2}(\beta |\kappa ^{\prime })\right]^{-1}$ & $H_1$ & 
  $=L_{x}^{2}+\kappa ^{\prime 2}L_{y}^{2}$\\
 & $\times\left[ -\kappa ^{\prime 2}\,{\rm sn}\,\alpha \,{\rm sn}
(\beta |\kappa ^{\prime }){\rm cn}(\beta |\kappa ^{\prime
  })\partial_\alpha -\,{\rm cn}\,\alpha \,{\rm {\rm dn}
}\,\alpha \,{\rm dn}(\beta |\kappa ^{\prime })
\partial_\beta \right]$ & & $=-\left[\kappa ^{2}{\rm 
{\rm cn}}^{2}\alpha +\kappa ^{\prime 2}{\rm cn}^{2}(\beta
|\kappa ^{\prime })\right]^{-1}\left( \kappa ^{\prime 2}{\rm sn}^{2}(\beta
|\kappa ^{\prime }){\rm \ }\partial_\alpha^{2}+{\rm \ 
{\rm dn}}\,^{2}\alpha\partial_\beta
^{2}\right) $\\
$L_y$ & $=i\left[\kappa ^{2}{\rm cn}^{2}\alpha +\kappa ^{\prime 2}
{\rm cn}^{2}(\beta |\kappa ^{\prime })\right]^{-1}$ & $H_2$ &
$=\kappa ^{2}L_{y}^{2}-L_{z}^{2}$\\
& $\times\left[ {\rm cn}\,\alpha \,{\rm sn}(\beta |\kappa
^{\prime }){\rm dn}(\beta |\kappa ^{\prime })
\partial_\alpha -\,{\rm sn}\,\alpha \,{\rm dn}\,\alpha 
{\rm cn}(\beta |\kappa ^{\prime })\partial_\beta 
\right]$ & & $=-\left[\kappa ^{2}{\rm {\rm cn
}}^{2}\alpha +\kappa ^{\prime 2}{\rm cn}^{2}(\beta |\kappa
^{\prime })\right]^{-1}\left( {\rm dn}^{2}(\beta |\kappa
^{\prime })\partial_\alpha^{2}
  +\kappa ^{2}{\rm sn}
^{2}\alpha \partial_\beta^{2}\right)$\\
$L_z$ & $=\left[\kappa ^{2}{\rm cn}^{2}\alpha +\kappa ^{\prime 2}
{\rm cn}^{2}(\beta |\kappa ^{\prime })\right]^{-1}$ & $H_3$ &
$=\kappa ^{2}L_{x}^{2}+\kappa ^{\prime 2}L_{z}^{2}$\\ 
 & $\times\left[ -{\rm dn}\,\alpha \,{\rm cn}(\beta |\kappa
^{\prime }){\rm dn}(\beta |\kappa ^{\prime })
\partial_\alpha -\kappa ^{2}\,{\rm sn}\,\alpha \,{\rm {\rm cn}
}\,\alpha \,{\rm sn}(\beta |\kappa ^{\prime })
\partial_\beta \right]$ & & $=\left[\kappa
^{2}{\rm cn}^{2}\alpha +\kappa ^{\prime 2}{\rm cn}
^{2}(\beta |\kappa ^{\prime })\right]^{-1}\left( \kappa ^{\prime 2}{\rm cn}
^{2}(\beta |\kappa ^{\prime })\;\partial_\alpha ^{2}
-\kappa ^{2}{\rm cn}^{2}\alpha\; \partial_\beta^{2}\right) $\\
& & &\\
$C_2$ & $= L_{z}^{2}-L_{x}^{2}-L_{y}^{2}$ & & \\
& $= \left[\kappa ^{2}{\rm cn}
^{2}\alpha +\kappa ^{\prime 2}{\rm cn}^{2}(\beta |\kappa ^{\prime
})\right]^{-1}\left( \partial_\alpha ^{2}+
\partial_\beta ^{2}\right) $ & & \\
& & &\\
\hline\hline
\end{tabular}
\endsideways
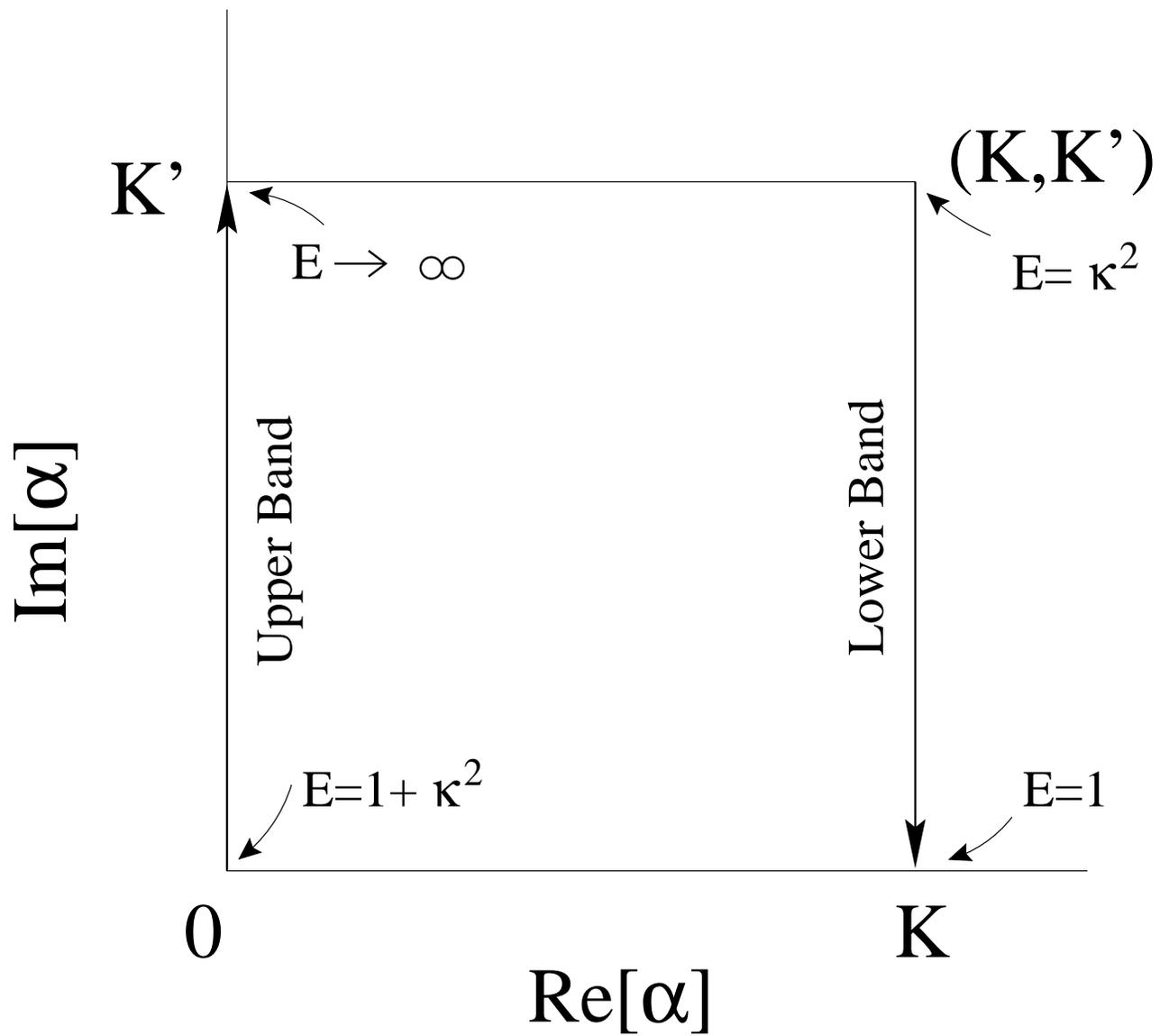
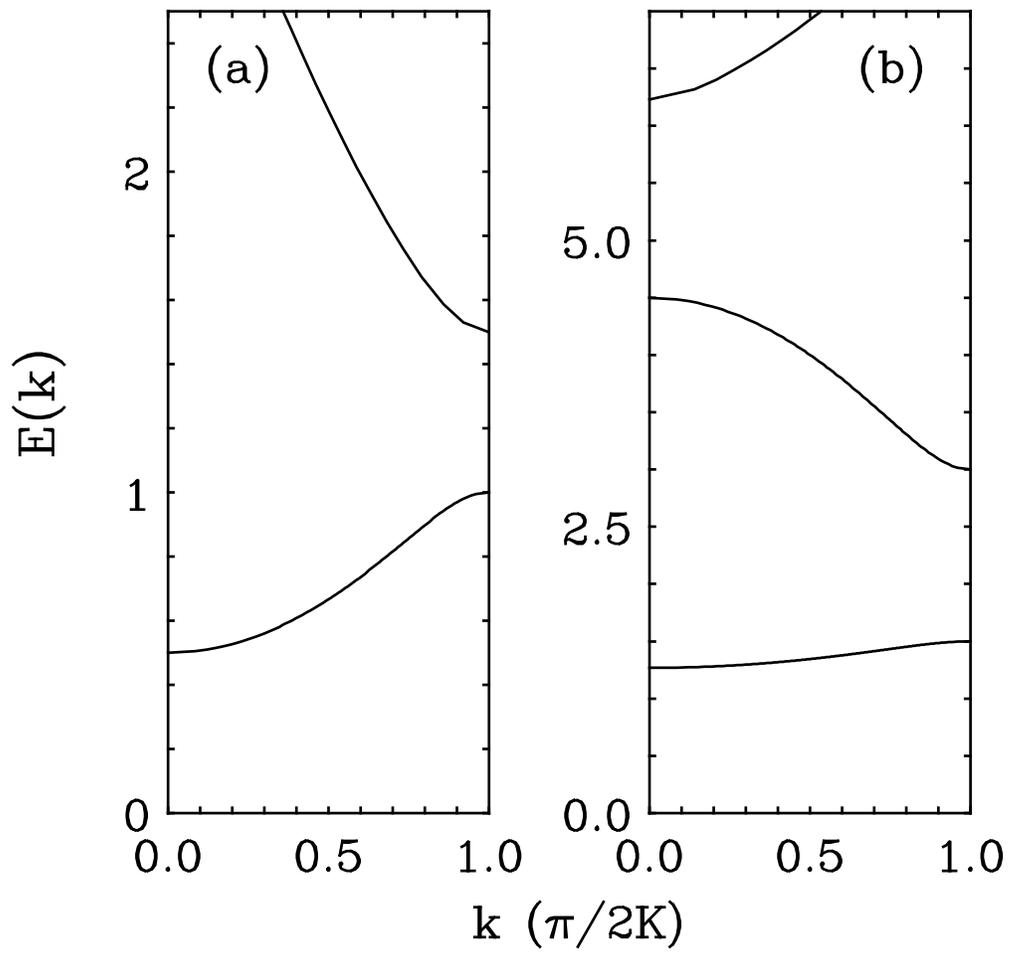
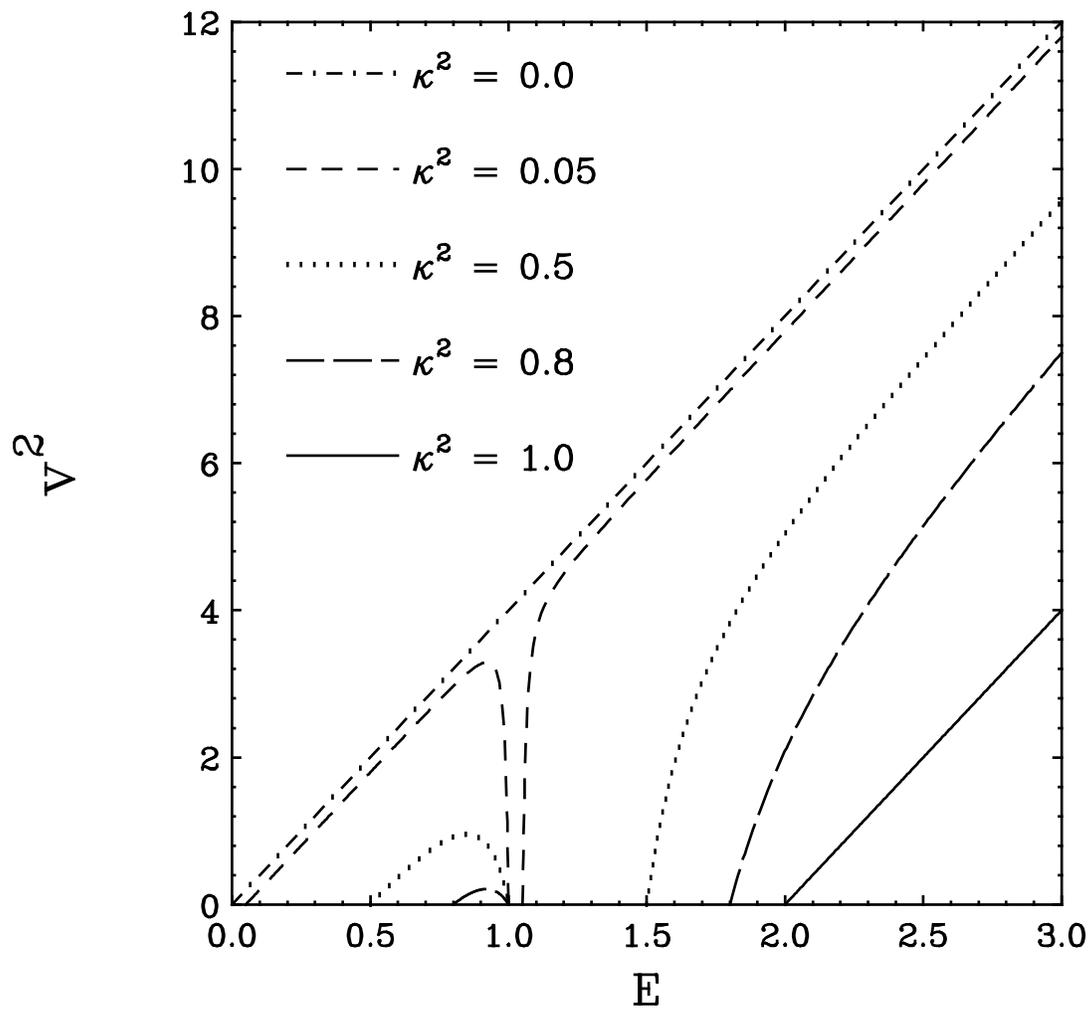
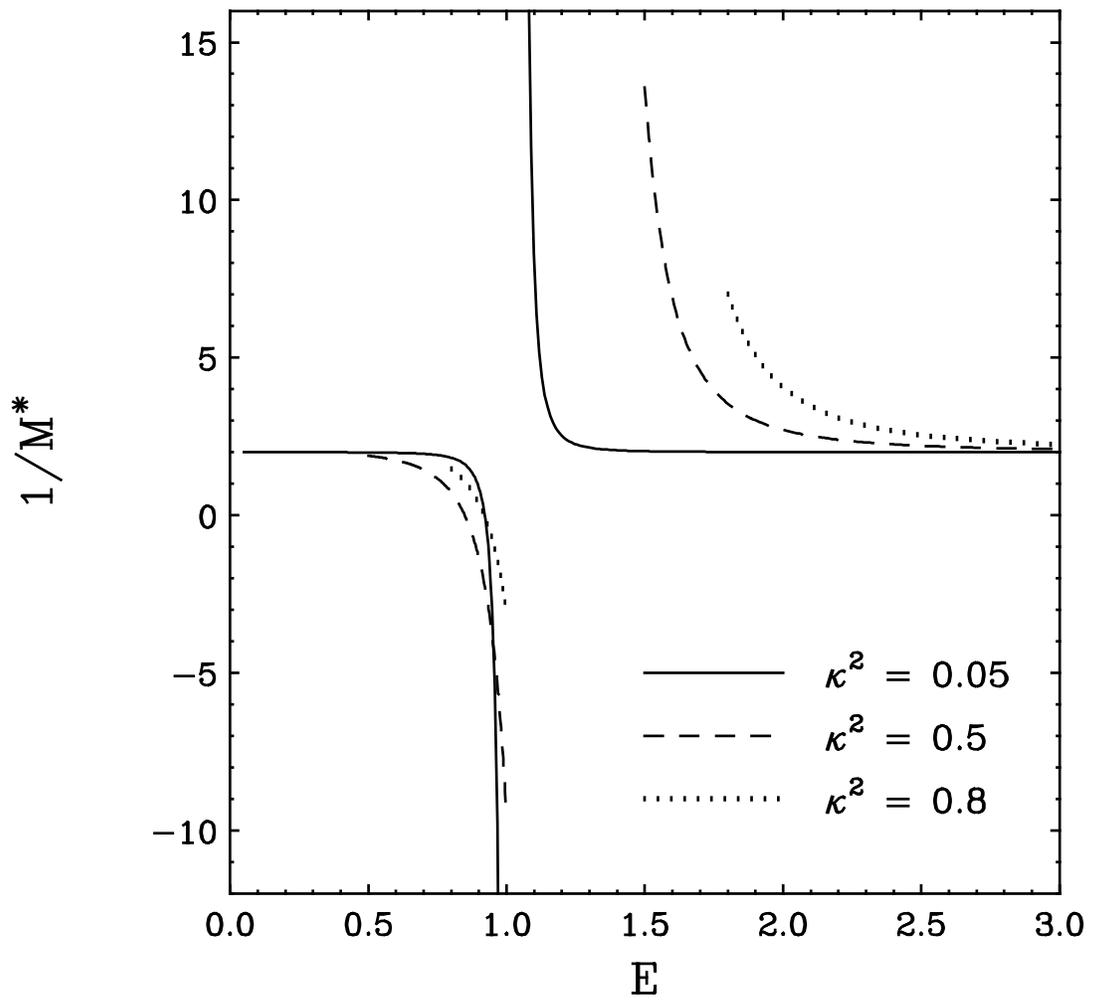
\end{document}